\documentclass{revtex4}
\usepackage{graphicx}
\usepackage{fancyhdr}
\usepackage{amsmath,amssymb}
\usepackage{color}

\begin{document}

\title{Two Higgs doublet models at future colliders}

%

\author{Koji Tsumura}
\affiliation{Department of Physics, Nagoya University, Nagoya 464-8602, Japan}

\begin{abstract}
The two Higgs doublet model (THDM) is a simple extension of the standard model, 
which can provide a low energy effective description of more fundamental theories. 
The model contains additional Higgs bosons, and predicts rich phenomenology 
especially due to the variation of Yukawa interactions. 
Under imposing a softly broken discrete symmetry, there are four independent types 
of Yukawa interactions in THDMs. 
In this review, we briefly summarize bounds from current experimental data on THDMs 
and implications at future collider experiments. 
We pay special attention to the collider phenomenology of the Type-X (lepton specific) 
THDM, and also discuss recent progress for $\tan\beta$ determination in THDMs. 
\end{abstract}

\maketitle

\thispagestyle{fancy}


\section{Introduction}
\vspace{-2ex}
A Higgs boson has been confirmed at the LHC~\cite{Ref:atlas,Ref:cms}. 
Clear peaks are observed in invariant mass distributions in $\gamma\gamma$ and 
$ZZ(\to 4\ell)$ decay channels, and excesses are also seen in several decay modes. 
It is consistent with the property of the Higgs boson in the standard model 
(SM)~\cite{Ref:atlas-comb,Ref:cms-comb}. 
In order to test the nature of the Higgs boson more accurately, 
precision measurement for the particle will be continued at the LHC. 
Expected uncertainties of the Higgs boson interaction strengths are evaluated 
in Ref.~\cite{Ref:ilc-Peskin}. 
At the International Linear Collider (ILC), the interaction strengths would be measured 
very precisely~\cite{Ref:ilc-Peskin,Ref:ilc-TDR}. 

The Higgs sector of the SM is constructed as the minimal form, i.e., the one Higgs doublet model. 
However, there is no fundamental reason to employ the minimal Higgs sector. 
The electroweak $\rho$ parameter has been measured very precisely, 
which seems to be a good guideline for constructing the extended Higgs sector. 
The measured value is very close to unity. 
The SM Higgs field, which develops the vacuum expectation value (VEV), is an $SU(2)_L$ 
doublet scalar with hypercharge $Y=1/2$. It predicts $\rho=1$ at the tree level, which 
is consistent with experimental data. 
Multi doublet extensions of the SM also hold $\rho=1$ at the tree level. 
If the electroweak symmetry is broken by the VEV of an $SU(2)$ triplet scalar with $Y=1$, 
$\rho=1/2$ is obtained, which is obviously disfavored by data. 
If a triplet develops  a VEV in addition to the VEV of the SM Higgs field, 
a VEV of the triplet is required to be very small.  
The next minimal representation, which keeps the $\rho$ parameter to be unity, 
is an $SU(2)$ septet with $Y=2$~\cite{Ref:7plet}. 
Therefore, the two Higgs doublet model (THDM) is a minimal viable extension of 
the SM Higgs boson sector. 

Multi doublet extensions of the SM Higgs sector have many variations 
due to the variation of Yukawa interactions. 
In general, there are two independent Yukawa interaction for each SM fermion. 
Thus, it leads to tree level flavor changing neutral currents (FCNCs), which 
are severely constrained by existing flavor data. 
In order to forbid the tree level FCNC, a discrete symmetry is introduced to THDMs~\cite{Ref:GW}. 
Under the discrete symmetry, there are four types of the Yukawa interactions~\cite{Ref:BHP,Ref:Grossman}. 
Summary of the flavor constraints and the collider phenomenology of each THDM are 
presented in Ref.~\cite{Ref:2hdm}.  

In order to explain many issues such as the naturalness problem of the Higgs boson mass, 
candidates for the cold dark matter, the origin of neutrino mass, etc..., 
the extended Higgs sector is often introduced in the beyond the SM. 
The most familiar THDM is the minimal supersymmetric standard model (MSSM), 
where the supersymmetry (SUSY) is designed to solve naturalness problem of 
the Higgs boson mass under the quantum correction. 
In SUSY models, the Higgs sector is automatically extended to have even numbers 
of the Higgs doublets due to the holomorohy of the superpotential. 
The Yukawa interactions is classified as the Type-II~\cite{Ref:Djouadi2,Ref:HHG,Ref:2hdm}. 
The THDM is also required in the gauged version of the seesaw model, where 
tiny neutrino masses are explained by the seesaw mechanism via triplet fermions~\cite{Ref:TypeX}. 
The model predicts the Type-X Yukawa interactions due to the anomaly cancellation.  
The Type-X THDMs are also used to explain 
the excess of the positron in the cosmic rays at PAMELA, Fermi, and Planck~\cite{Ref:Goh}, 
the tiny neutrino mass by a three-loop radiative seesaw model~\cite{Ref:AKS}, 
and the deviation in the muon anomalous magnetic moment~\cite{Ref:g-2}. 

In this talk, we summarize flavor constraints and direct search bounds on the THDMs. 
%
In the Type-X (lepton specific) THDM, relatively light non-standard Higgs bosons are 
allowed after considering all experimental data. 
In such a scenario, we can access to the new Higgs bosons at future colliders. 
The search strategies and the mass determinations of the additional Higgs bosons 
are presented. 
In THDMs, the ratio of the VEVs for two doublets, $\tan\beta$, is an important parameter. 
The information about $\tan\beta$ can be extracted from not only non-standard Higgs bosons 
but also SM-like Higgs boson phenomenology. 
We discuss recent studies of $\tan\beta$ determination methods at the ILC.

\section{THDM$\text{s}$ confront experimental data}
\vspace{-1ex}
We here consider the Higgs potential which consists of two Higgs doublet fields $\Phi_a (a=1, 2)$ as 
\begin{align}
{\mathcal V}^\text{THDM} 
&= +m_1^2\Phi_1^\dag\Phi_1+m_2^2\Phi_2^\dag\Phi_2
-m_3^2\left(\Phi_1^\dag\Phi_2+\Phi_2^\dag\Phi_1\right)
+\frac{\lambda_1}2(\Phi_1^\dag\Phi_1)^2
+\frac{\lambda_2}2(\Phi_2^\dag\Phi_2)^2\nonumber \\
&\qquad+\lambda_3(\Phi_1^\dag\Phi_1)(\Phi_2^\dag\Phi_2)
+\lambda_4(\Phi_1^\dag\Phi_2)(\Phi_2^\dag\Phi_1)
+\frac{\lambda_5}2\left[(\Phi_1^\dag\Phi_2)^2
+(\Phi_2^\dag\Phi_1)^2\right], \label{Eq:HiggsPot}
\end{align}
where a softly broken discrete symmetry is imposed~\cite{Ref:GW}. 
The component fields are parameterized as 
\begin{align}
\Phi_i=\begin{pmatrix}i\,\omega_i^+\\\frac1{\sqrt2}(v_i+h_i-i\,z_i)
\end{pmatrix}.
\end{align}
Assuming the CP-invariant Higgs sector mass eigenstates are defined by the following rotations as  
\begin{align}
\begin{pmatrix}h_1\\h_2\end{pmatrix}=\text{R}(\alpha)
\begin{pmatrix}H\\h\end{pmatrix},\quad
\begin{pmatrix}z_1\\z_2\end{pmatrix}=\text{R}(\beta)
\begin{pmatrix}z\\A\end{pmatrix},\quad
\begin{pmatrix}\omega_1^+\\\omega_2^+\end{pmatrix}=\text{R}(\beta)
\begin{pmatrix}\omega^+\\H^+\end{pmatrix},
\end{align}
where $h$ and $H$ are CP even states, $A$ is a CP odd state, $H^\pm$ are charged states, 
$z$ and $\omega^\pm$ are Nambu-Goldstone bosons, and 
\begin{align}
\text{R}(\theta)=\begin{pmatrix}\cos\theta&-\sin\theta \\
\sin\theta&\cos\theta\end{pmatrix}.
\end{align}
\begin{table}[tb]
\begin{center}
\begin{tabular}{|c||c|c|c|c|c|c|}
\hline & $\Phi_1$ & $\Phi_2$ & $u_R^{}$ & $d_R^{}$ & $\ell_R^{}$ &
 $Q_L$, $L_L$ \\  \hline
Type-I  & $+$ & $-$ & $-$ & $-$ & $-$ & $+$ \\
Type-II & $+$ & $-$ & $-$ & $+$ & $+$ & $+$ \\
Type-X  & $+$ & $-$ & $-$ & $-$ & $+$ & $+$ \\
Type-Y  & $+$ & $-$ & $-$ & $+$ & $-$ & $+$ \\
\hline
\end{tabular}
\end{center}
\caption{Parity assignments under the softly broken $Z_2$ symmetry~\cite{Ref:AKTY}.}
 \label{Tab:type}
\end{table}
In order to forbid tree level FCNCs the discrete symmetry is introduced, 
where the parity assignments for each field are listed in TABLE.~\ref{Tab:type}. 
There are four independent combinations of parity assignments. 
Using the component fields, Yukawa interactions are given by 
\begin{align}
{\mathcal L}_\text{yukawa}^\text{THDM} =
&-\sum_{f=u,d,\ell} \Bigl[
+\frac{m_f}{v}\, \xi_h^f\, {\overline f}fh
+\frac{m_f}{v}\, \xi_H^f\, {\overline f}fH
-i\frac{m_f}{v}\, \xi_A^f\, {\overline f}\gamma_5fA
\Bigr] \nonumber\\
&-\Bigl\{ +\frac{\sqrt2V_{ud}}{v}\, \overline{u}
\bigl[ +m_u\, \xi_A^u\, \text{P}_L+m_d\, \xi_A^d\, \text{P}_R\bigr]d\,H^+
+\frac{\sqrt2\, m_\ell\, \xi_A^\ell}{v}\overline{\nu_L^{}}\ell_R^{}H^+
+\text{H.c.} \Bigr\},\label{Eq:Yukawa}
\end{align}
where the scaling factors $\xi^\phi_f (\phi=h, H, A)$ are determined 
by the Higgs mixing parameters $\alpha$ and $\beta$ in TABLE.~\ref{Tab:ScalingFactor}. 
\begin{table}[tb]
\begin{center}
\begin{tabular}{|c||c|c|c|c|c|c|c|c|c|}
\hline
& $\xi_h^u$ & $\xi_h^d$ & $\xi_h^\ell$
& $\xi_H^u$ & $\xi_H^d$ & $\xi_H^\ell$
& $\xi_A^u$ & $\xi_A^d$ & $\xi_A^\ell$ \\ \hline
Type-I
& $c_\alpha/s_\beta$ & $c_\alpha/s_\beta$ & $c_\alpha/s_\beta$
& $s_\alpha/s_\beta$ & $s_\alpha/s_\beta$ & $s_\alpha/s_\beta$
& $\cot\beta$ & $-\cot\beta$ & $-\cot\beta$ \\
Type-II
& $c_\alpha/s_\beta$ & $-s_\alpha/c_\beta$ & $-s_\alpha/c_\beta$
& $s_\alpha/s_\beta$ & $c_\alpha/c_\beta$ & $c_\alpha/c_\beta$
& $\cot\beta$ & $\tan\beta$ & $\tan\beta$ \\
Type-X
& $c_\alpha/s_\beta$ & $c_\alpha/s_\beta$ & $-s_\alpha/c_\beta$
& $s_\alpha/s_\beta$ & $s_\alpha/s_\beta$ & $c_\alpha/c_\beta$
& $\cot\beta$ & $-\cot\beta$ & $\tan\beta$ \\
Type-Y
& $c_\alpha/s_\beta$ & $-s_\alpha/c_\beta$ & $c_\alpha/s_\beta$
& $s_\alpha/s_\beta$ & $c_\alpha/c_\beta$ & $s_\alpha/s_\beta$
& $\cot\beta$ & $\tan\beta$ & $-\cot\beta$ \\
\hline
\end{tabular}
\end{center}
\caption{The scaling factors in each type of Yukawa interactions in
 Eq.~\eqref{Eq:Yukawa}~\cite{Ref:AKTY}.} \label{Tab:ScalingFactor}
\end{table}
%

The various constraints on the THDMs have been studied in the literature. 
From the theoretical consideration such as perturbativitive unitarity~\cite{Ref:PU-2hdm} 
and the vacuum stability~\cite{Ref:VS-2hdm}, parameters in the Higgs potential are bounded. 
Too large mass splittings among additional Higgs bosons are also constrained 
because mass differences are induced from the violation of the custodial symmetry 
in the two Higgs doublet model potential~\cite{Ref:rho-2hdm,Ref:rho2-2hdm,Ref:rho3-2hdm,Ref:KOTT}. 
These above bounds are originated from the Higgs potential, and hence these are independent 
of types of Yukawa interactions. 

The Yukawa interactions of THDMs receive severe constraints from flavor physics experiment, 
because interaction strengths of the additional Higgs bosons with fermions are 
enhanced/suppressed depending on the type of Yukawa interactions as shown in TABLE.~\ref{Tab:ScalingFactor}. 
In the Type-I THDM, Yukawa interactions of $h$, $H$ and $A$ are universally corrected 
for all fermions. For small $\tan\beta$, the mass of charged Higgs boson is bounded by 
$Z\to b\bar b$~\cite{Ref:HL}, and $B \to X_s \gamma$~\cite{Ref:BHP,Ref:bsg}. 
With nearly SM-like condition $\sin(\beta-\alpha) \simeq 1$, $H$ and $A$ become 
fermiophobic for large $\tan\beta$. 
Thus, all flavor constraints on the Type-I THDM can be evaded for moderate or large 
$\tan\beta$ even with relatively light additional Higgs bosons. 
In the Type-II THDM, the stringent bound on the mass of $H^\pm$ comes from 
$B \to X_s \gamma$~\cite{Ref:BHP}. For the wide range of  $\tan\beta~(\gtrsim 2)$, 
$m_{H^\pm}^{} \gtrsim 300$ GeV is obtained~\cite{Ref:bsg}. For large $\tan\beta~(\gtrsim 40)$, 
the more stronger bound is derived from $B \to \tau \nu$~\cite{Ref:btaunu}. 
Note that the bound from $B \to X_s \gamma$ may be weakened due to the cancellations 
in the loop diagram, i.e., the stop-chargino loop diagram in the MSSM~\cite{Ref:GO}.
In the Type-X THDM, there are no severe constraint from $B$ decay data due to 
the less interaction with quarks. Purely leptonic observables such as leptonic decays of tau lepton 
can constrain light charged Higgs boson with very large $\tan\beta$~\cite{Ref:tau}. 
In the Type-Y THDM, Yukawa interactions with quarks are the same as in the Type-II THDM. 
Thus, severe constraint is also obtained from $B \to X_s \gamma$. 

The additional Higgs bosons have been searched at the LEP~\cite{Ref:LEP,Ref:LEP2}. 
For $HA$ pair production, their $b\bar b$, $\tau^+\tau^-$ decay modes are analyzed 
to obtain the lower mass bound~\cite{Ref:LEP}. 
For charged Higgs pair production, their $c\bar s (s \bar c)$, $\tau\nu$ decay channels are 
used~\cite{Ref:LEP2}. 
These production cross section are mostly independent of $\tan\beta$ in the nearly 
SM-like limit. Therefore, the lower mass bounds of about one hundred GeV are obtained 
almost independently from types of Yukawa interaction. 
If there are relatively large mass splittings, $\phi \to \phi' V (V=W,Z)$ decay modes 
may become important, where $\phi$ denotes an additional Higgs boson. 
Since the production cross sections for additional Higgs bosons are small at the LHC
due to the electroweak interaction, possible production mechanism would be the gluon fusion 
to $H/A$, $H/A$ radiations from quark pairs, and the top quark decays into charged Higgs 
bosons via the enhanced Yukawa interactions. 
In the Type-I THDM, the small $\tan\beta$ region is only accessible parameter space 
through these above processes because of fermiophobic nature of additional Higgs bosons. 
In the Type-II THDM, $H/A$ production in association with $b\bar b$ becomes significant  
for large $\tan\beta$ due to the enhanced Yukawa interaction with bottom quarks. 
The experimental bounds are obtained in the context of the MSSM, 
where the lower bound of $\tan\beta$ is not applicable for the Type-II THDM 
since it comes from the SUSY specific property~\cite{Ref:mssm-HA-atlas,Ref:mssm-HA-cms}. 
In general, small $\tan\beta$ and light $H/A$ are ruled out. 
For instance, $m_A^{} = 250$ GeV is excluded for $\tan\beta > 5$~\cite{Ref:mssm-HA-atlas}.
From non-observation of $t \to H^+ b$ decay, small $\tan\beta$ is also ruled out~\cite{Ref:mssm-H+}. 
In the Type-X THDM, $H/A$ is difficult to be produced via the Yukawa interaction
because only leptonic Yukawa interactions are enhanced for large $\tan\beta$. 
Therefore, there is no severe bound from the direct search except for LEP bound~\cite{Ref:AKTY}. 
In the Type-Y THDM, additional Higgs bosons can be produced similarly to those in the Type-II THDM. 
Since the leptonic Yukawa interactions are suppressed for $\tan\beta$, produced Higgs bosons tend
to decay hadronically. Thus, it is difficult to be observed in the same process as in the Type-II THDM.

\section{Implications at future colliders}
\vspace{-1ex}
We here focus on the following two subjects. 
In subsection A, collider phenomenology of relatively light additional Higgs bosons 
in the Type-X THDM is discussed. 
In subsection B, sensitivities of $\tan\beta$ in THDMs at the ILC are studied by several methods. 
\vspace{-1em}

\subsection{Direct search for $H/A$ in Type-X THDM}
\vspace{-1ex}
Since masses of the additional Higgs bosons can be of the order of one hundred GeV
in the Type-X THDM, these particles can be produced at the LHC and also at the ILC. 
In this subsection, we introduce search strategies of non-standard Higgs bosons in each experiment. 

At the LHC, relatively light additional Higgs bosons can be pair produced as $q\bar q \to HA$. 
Generated $H/A$ subsequently decays into $\tau^+\tau^-$ for $\tan\beta \gtrsim 3$ 
in the Type-X THDM~\cite{Ref:AKTY}. 
Therefore, multi tau lepton events are the distinctive signal of this model. 
In Ref.~\cite{Ref:KTY-LHC}, we made detailed simulation studies for such multi tau lepton signatures. 
One of the key selection cuts is a requirement of the high multiplicity of tau-jets, 
where a tau-jet candidate is identified by a jet which contains 1 or 3 charged hadrons 
in a small cone ($R<0.15$). 
As for the $HA \to 4\tau$ signal, background events are well reduced by requiring two or 
more tau-jets. 
For example, in the $4\tau_h$ channel, where $\tau_h$ is a hadronically tagged tau lepton, 
the signal significance $S=\sqrt{2[(s+b)\ln(1+s/b)-s]}$ is expected to be 9.3 
after the selection cuts, where $m_H^{}=130$ GeV, $m_A^{}=170$ GeV, $\sqrt{s}=14$ TeV and 
$L=100 \text{fb}^{-1}$ are assumed~\cite{Ref:KTY-LHC}. 
Other tau lepton decay channels ($3\tau_h \ell, 2\tau 2\ell$) are also useful to see multi tau lepton 
signatures, where $\ell = e, \mu$~\cite{Ref:KTY-LHC}.  Thus, the model would be easily test at the LHC. 
In the more recent study, the same signed dilepton with hadronically tagged tau leptons are also 
discussed as a promising signal of the Type-X THDM~\cite{Ref:LSWY}. 
For the mass determination of $H/A$, the dimuon decay channel of $H$ or $A$ seems to be useful. 
Assuming one muonically decaying non-standard Higgs boson, we expect $2\mu2\tau$ signatures, 
which are only 0.7 \% of $HA$ pair production events. 
This decay chain can be seen due to the strong suppression of background 
events using the sharp invariant mass cut of dimuons if we have large enough luminosity. 
The $\tau^+\tau^-$ invariant mass can also be constructed in this channels by the help of 
the collinear approximation. Thus, $HA$ pair production can be explored at the LHC. 
For more details and also for the charged Higgs productions $q\bar q' \to W^* \to \phi H^\pm$ 
and $q\bar q' \to Z^* \to H^+ H^-$, see Ref.~\cite{Ref:KTY-LHC}. 
Note that the former process is possible only at hadron colliders, 
which is useful to analyze charged Higgs bosons at the LHC~\cite{Ref:KTY-LHC}.

At the ILC, the non-standard Higgs bosons can be pair produced in $e^+e^-\to HA$ and 
$e^+e^-\to H^+H^-$, if they are light enough. 
Similar decay chains discussed in the LHC case can be analyzed for $HA$ pair production. 
Thanks to small background event rates, the signal excess can be easily seen by more than 
$13\sigma$ level in $4\tau_h$ channel only with $10~\text{fb}^{-1}$, where the collision energy 
is assumed to be 500 GeV, and $m_H^{}=130$ GeV, $m_A^{}=170$ GeV~\cite{Ref:KTY-ILC}. 
Although there are at least four missing neutrinos, momenta for all the tau leptons can be 
reconstructed using the collinear approximation. 
In the LHC case, we rely only on the conservation of the transverse momentum, and hence 
only two missing momenta are allowed to reconstruct tau lepton momenta. 
On the other hand, the initial four momentum is known at the ILC so that four missing momenta 
from $4\tau$ events are fully reconstructed. Therefore, precise mass determination and also 
the test of pair production can be easily performed at the ILC through the $HA\to 4\tau$ decay 
chain. 
The detailed information is given in Ref.~\cite{Ref:KTY-ILC}.
\vspace{-1ex}

\subsection{$\tan\beta$ determination at the ILC}
\vspace{-1ex}
The ratio of VEVs, namely $\tan\beta=v_2/v_1$ is an important parameter in THDMs. 
The $\tan\beta$ measurement in the MSSM has been investigated in Refs.~\cite{Ref:TanB1,Ref:TanB2}. 
The branching ratios of $H/A$ into $b\bar b$ are dependent on $\tan\beta$, and reach 
a saturation point, which is about 90\% for moderate $\tan\beta$ values 
(10 \% for $\tau^+\tau^-$ decay mode). 
Thus, the branching ratio measurement of $H/A$ is sensitive only for small $\tan\beta$~\cite{Ref:TanB1,Ref:TanB2}. 
While for large $\tan\beta$, total decay widths of $H/A$ are roughly proportional to 
the square of $\tan\beta$. 
If the total widths are wider than the detector resolution, information of $\tan\beta$ 
can be extracted~\cite{Ref:TanB1,Ref:TanB2}. 
For both methods to determine $\tan\beta$, $HA$ pair production with $e^+e^-$ collision 
energy of 500 GeV is assumed. Due to the limitation of collision energy, $m_A^{}$ is taken 
to be light, i.e., 200 GeV~\cite{Ref:TanB1,Ref:TanB2}.
However, the recent LHC data exclude the most of the parameter space in the MSSM with 
$m_A^{}=200$ GeV except for $\tan\beta <5$.

Since $y_b/y_b^\text{SM}=\sin(\beta-\alpha)-\tan\beta\cos(\beta-\alpha)$, where $y_b$ is 
the Yukawa coupling strength for $h$ with bottom quarks in the Type-II THDM while 
$y_b^\text{SM}$ is that in the SM, the deviation becomes significant for large $\tan\beta$. 
Information of $\tan\beta$ can be extracted from the SM-like Higgs boson decay 
if we observe deviations in the SM-like Higgs gauge coupling, $\kappa_V^{} =\sin(\beta-\alpha) \ne 1$.
In the MSSM, $\sin(\beta-\alpha)$ is restricted to be very close to unity, because the SUSY 
requires the relation, $\sin(\beta-\alpha) \simeq 1 - 2m_Z^4/(m_A^4 \tan^2\beta)$ 
for large $\tan\beta$ with $m_Z^{} \ll m_A^{}$.
On the other hand, interactions of the SM-like Higgs boson $h$ in the Type-II THDM 
can be different from those in the MSSM. The Higgs mixing parameters $\alpha$ and 
$\beta$ are, in general, independent in the Type-II THDM. 
Thus, precise measurement of $\sin(\beta-\alpha)$ is crucial to independently determine $\tan\beta$. 

At the ILC with $\sqrt{s}=250$ GeV and $L=250~\text{fb}^{-1}$, the correction factor to 
the gauge coupling of the SM-like Higgs boson is expected to be measured by the
$\Delta \sigma_{Zh}^{}/\sigma_{Zh} = 2.5 \%$ accuracy, where the leptonic decays of 
the recoil $Z$ boson are assumed in $e^+e^-\to Zh$~\cite{Ref:Zh-lep}. The uncertainty 
would be reduced to 0.8 \% by taking into account hadronic decay modes~\cite{Ref:Zh-had}. 
On the other hand, the cross section times the decay branching ratio of the SM-like Higgs boson into 
$b\bar b$ could be measured more precisely as 
$\Delta (\sigma_{Zh}^{}{\mathcal B}_{bb}^h)/(\sigma_{Zh}^{}{\mathcal B}_{bb}^h) = 1 \%$~\cite{Ref:h-BR}. 
Thus, expected uncertainty for the independent determination of ${\mathcal B}_{bb}^h$ is 
$\Delta{\mathcal B}_{bb}^h/{\mathcal B}_{bb}^h =2.7 \%$ at $2 \sigma$ level 
(assuming leptonic decay channels of the recoil $Z$ boson).

\begin{figure}[tb]
 \centering
 \includegraphics[height=5.2cm]{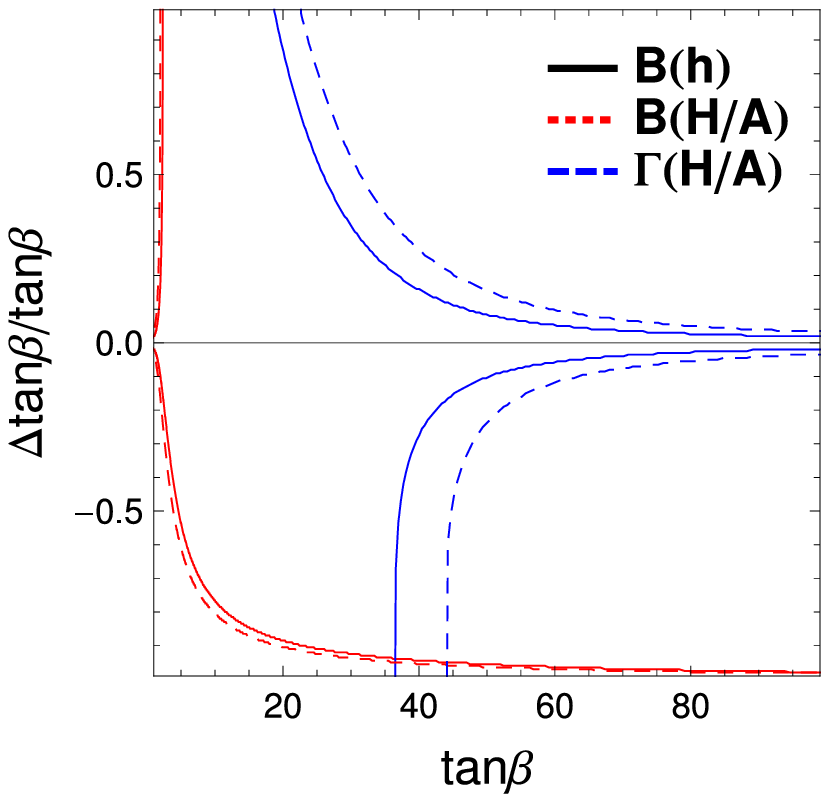} 
 \includegraphics[height=5.2cm]{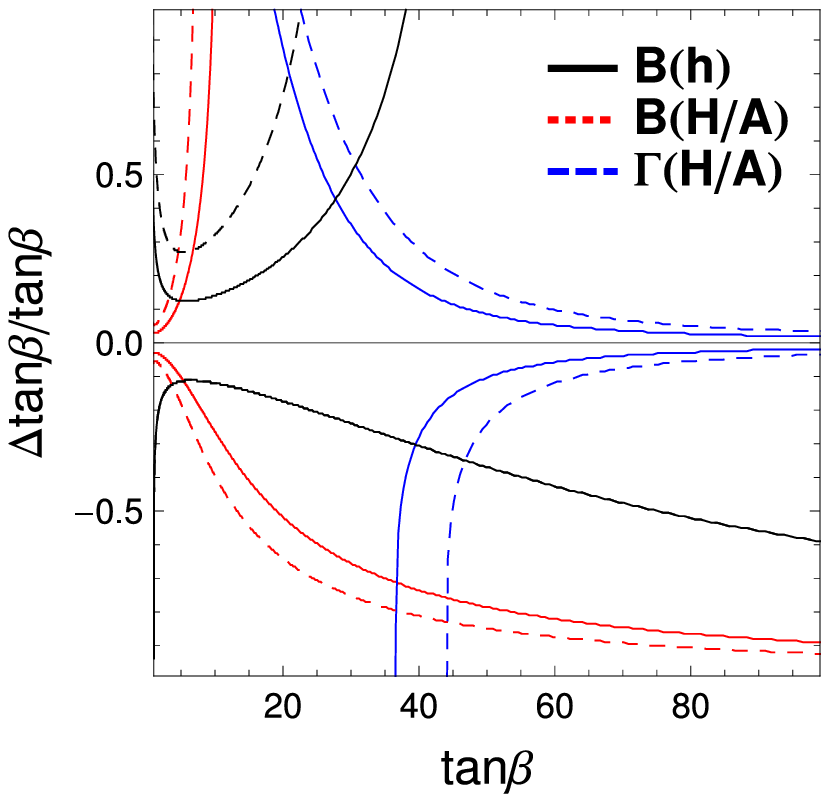} 
 \includegraphics[height=5.2cm]{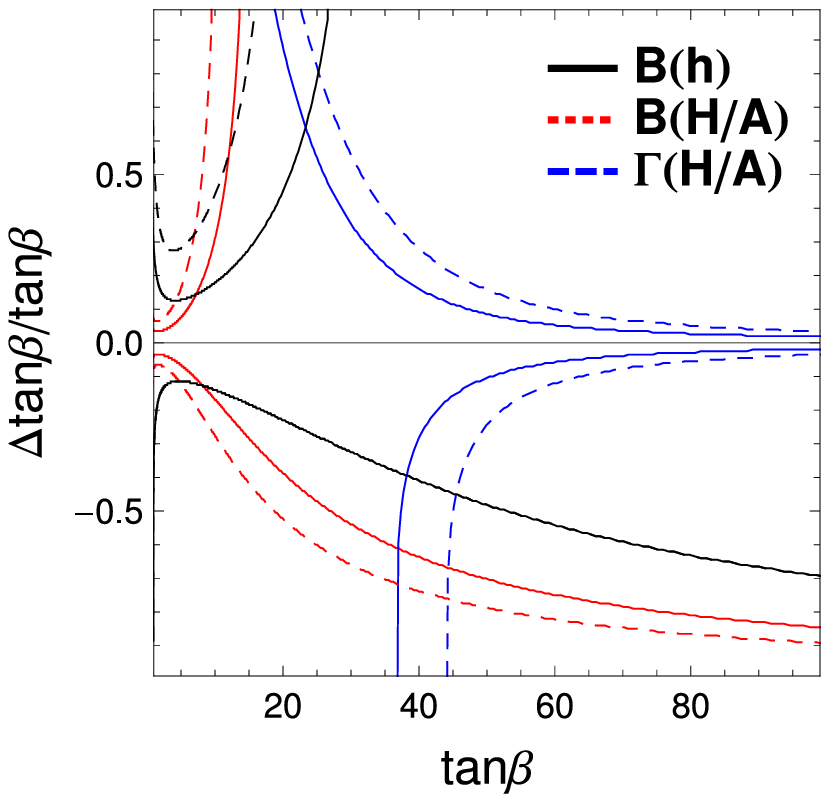} 
 \caption{Sensitivities of $\tan\beta$ by several methods in the Type-II THDM~\cite{Ref:KTY-TanB}. 
 From the left to right, $\sin^2(\beta-\alpha)$ is taken to be $1$, $0.99$, and $0.98$, with 
 $\cos(\beta-\alpha) \le 0$. 
 The branching ratio measurement method of $H/A\to b\bar b$ (red curves), 
 the total width measurement method of $H/A$ (blue curves), and 
 the branching ratio measurement method of $h\to b\bar b$ (black curves) are shown. 
For $HA$ production, $\sqrt{500}$ GeV and $L=250~\text{fb}^{-1}$ is assumed 
with $m_H^{}=m_A^{}=200$ GeV. 
For the $h\to b\bar b$ measurement, 
$\Delta{\mathcal B}/{\mathcal B} = 1.3~\% (1\sigma)$ and $2.7~\% (2 \sigma)$ 
are chosen {\it without} specifying masses of additional scalar bosons. 
 The dashed curves stand for the $2~\sigma$ sensitivities. 
 }
 \label{FIG:2HDM-II}
\end{figure}

In FIG.~\ref{FIG:2HDM-II}, the $1 \sigma$ ($2 \sigma$) sensitivities of $\tan\beta$ 
by several methods are shown in the solid (dashed) curves~\cite{Ref:KTY-TanB}. 
The red, blue, and black curves denote the method by 
i) the branching ratio measurement of $H/A$, 
ii) the total width measurement of $H/A$, and 
iii) the branching ratio measurement of $h$, respectively. 
The $1\sigma$ sensitivities for each method are defined; 
\begin{enumerate}
\def\theenumi{\roman{enumi})}
\item $N(\tan\beta\pm\Delta\tan\beta)=N(\tan\beta) \pm \sqrt{N(\tan\beta)}$, 
where $N(\tan\beta)$ is the number of $4b$ events from $HA$ production 
after the selection cuts. The acceptance of the $4b$ final states  
is evaluated to be 51~\%. 
\item $\Gamma^R_{H/A}(\tan\beta \pm \Delta\tan\beta)
=\Gamma^R_{H/A}(\tan\beta) \pm \Delta\Gamma^R_{H/A}(\tan\beta)$, 
where the observable averaged width is 
$\Gamma^R_{H/A} = \frac12[\sqrt{(\Gamma_\text{tot}^H)^2+(\Gamma_\text{res})^2}
+\sqrt{(\Gamma_\text{tot}^A)^2+(\Gamma_\text{res})^2}]$, and $1\sigma$ error is 
$\Delta\Gamma^R_{H/A}(\tan\beta)=[(\Gamma^R_{H/A}/\sqrt{2N(\tan\beta)})^2
+(\Delta\Gamma_\text{res}^\text{sys})^2]^{1/2}$. 
The detector resolution is evaluated to be $\Gamma_\text{res} = 11.3$~GeV, 
and 10 \% systematic error is assumed. 
$N(\tan\beta)$ is the number of events after the acceptance cuts with 
mass window cut of $M_{bb} \pm 10$ GeV, where the selection efficiency of 
mass window cut is estimated to be 42~\%.
\item ${\mathcal B}^h_{bb}(\tan\beta \pm \Delta\tan\beta) 
= {\mathcal B}^h_{bb}(\tan\beta) \pm \Delta{\mathcal B}^h_{bb}(\tan\beta)$.
\end{enumerate}
%
The condition, $\cos(\beta-\alpha) \le 0$, is taken, which is the same sign as in the MSSM. 
Note that $\cos(\beta-\alpha)$ can be positive in general THDMs. 
The results for $\cos(\beta-\alpha) \ge 0$ and detailed studies are given in 
Ref.~\cite{Ref:KTY-TanB}.  
The masses of $H/A$ are taken to be 200 GeV for $HA$ pair production, 
which is however ruled out for large $\tan\beta$ by the LHC data. 
On the other hand, for the SM-like Higgs boson study, masses of $H/A$ are not specified. 
Therefore, the sensitivity of $\tan\beta$ by the SM-like Higgs boson decay 
are still applicable in the Type-II THDM. 
Note that there is no black curve in the SM-like limit $\sin(\beta-\alpha)=1$ (left panel), 
because tree level couplings of the SM-like Higgs boson with fermions and weak gauge bosons 
are the same as those of the SM one. 
For the middle and the right panels, the sensitivities by the SM-like Higgs boson decay become 
rapidly worse for large $\tan\beta$, where ${\mathcal B}^h_{bb}$ is saturated. 
Such large corrections of the SM-like Higgs boson decay should be excluded by the LHC data. 

In the Type-X THDM, leptonic Yukawa interactions of $H/A$ are enhanced by $\tan\beta$ 
with $\sin(\beta-\alpha)\simeq 1$. Thus, the decay of $H/A$ into $\tau^+\tau^-$ becomes 
dominant if $\tan\beta > 3$. For the wide range of the parameter space, $4\tau$ final states are 
expected following $HA$ pair production. Therefore, measurements of $H/A \to \tau^+\tau^-$ 
can probe $\tan\beta$. The correction to Yukawa interactions of $h$ with tau leptons is also 
useful to explore $\tan\beta$. 

\begin{figure}[tb]
 \centering
 \includegraphics[height=5.2cm]{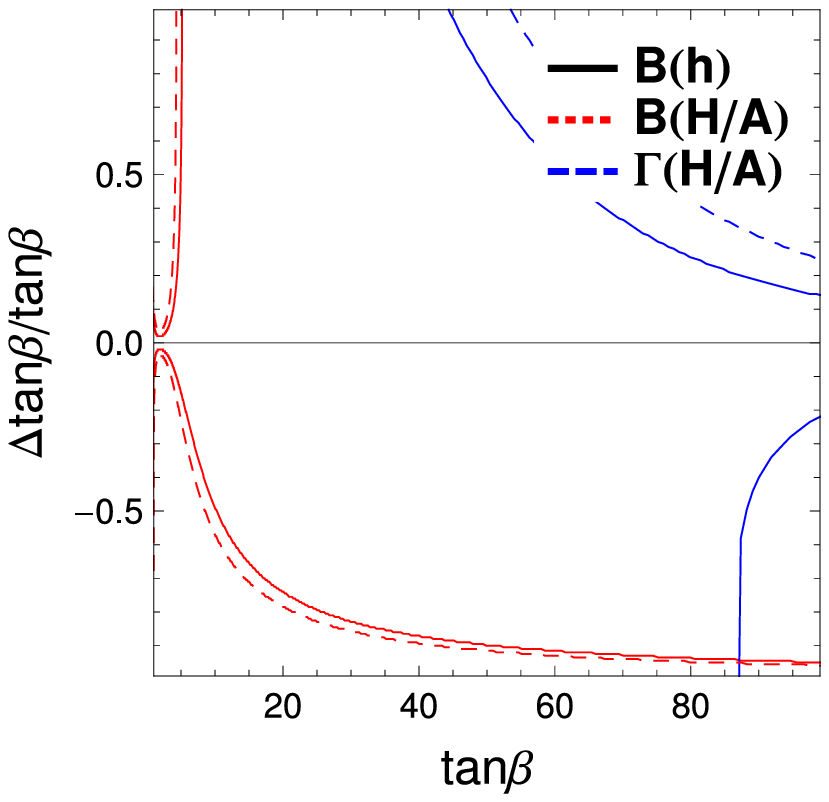} 
 \includegraphics[height=5.2cm]{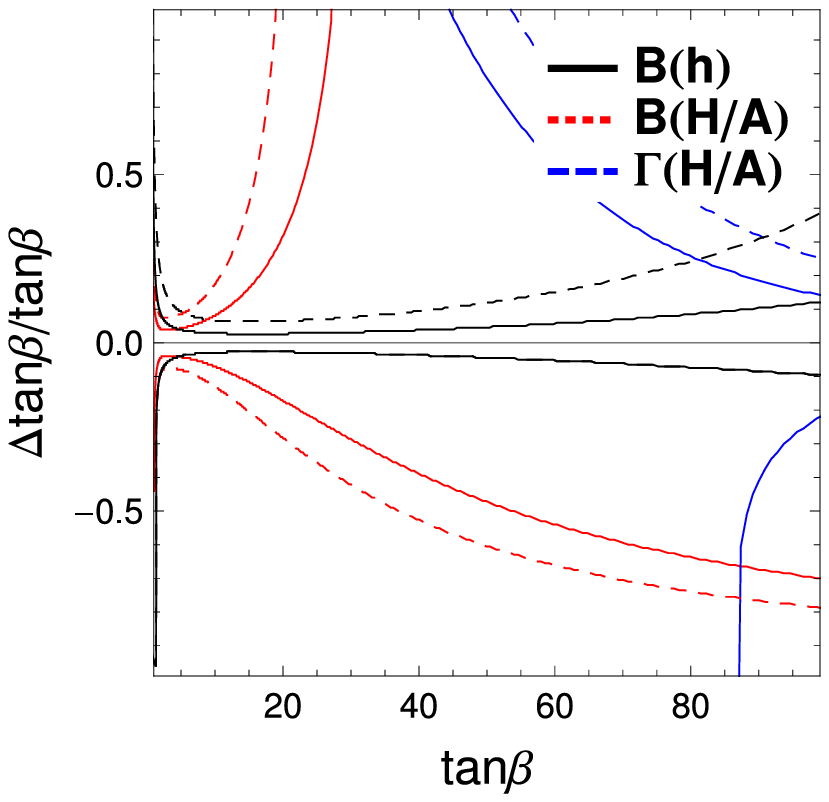} 
 \includegraphics[height=5.2cm]{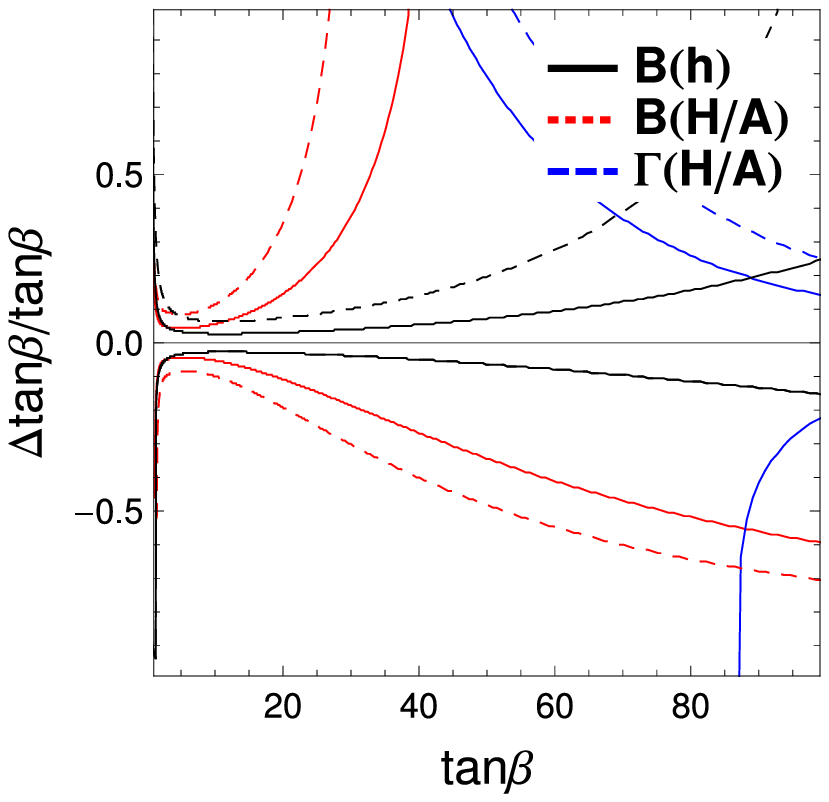} 
 \caption{Sensitivities of $\tan\beta$ by several methods in the Type-X THDM~\cite{Ref:KTY-TanB}. 
 Figures are set by the similar manner with FIG.~\ref{FIG:2HDM-II}.
For the $h\to \tau^+ \tau^-$ measurement, 
$\Delta{\mathcal B}/{\mathcal B} = 2~\% (1\sigma)$ and $5~\% (2\sigma)$ 
are chosen. 
 }
 \label{FIG:2HDM-X}
\end{figure}

In FIG.~\ref{FIG:2HDM-X}, sensitivities of $\tan\beta$ in the Type-X THDM are 
presented~\cite{Ref:KTY-TanB}. 
The model parameters are chosen similarly to as in FIG.~\ref{FIG:2HDM-II}. 
Instead of $b\bar b$ decay channels oin the Type-II THDM, 
$\tau^+\tau^-$ decay modes are used to determine $\tan\beta$ in the Type-X THDM. 
For $HA$ production, the acceptance for $4\tau$ events is estimated to be 47 \%. 
The detector resolution of $\tau^+\tau^-$ invariant mass is evaluated as 6.8 GeV, 
where the selection efficiency of $M_{\tau\tau}\pm 10$ GeV cut is 31 \%. 
For the measurement of the SM-like Higgs boson decay, the expected uncertainty is
$\Delta {\mathcal B}^h_{\tau\tau}/{\mathcal B}^h_{\tau\tau} = 5~\%$ at $2 \sigma$. 
For the wide range of the parameter region, the precision measurement of 
the branching ratio for $h$ gives the best sensitivity of $\tan\beta$. 
Since relatively light additional Higgs bosons are allowed in the Type-X THDM, 
direct measurements of $H/A$ can also probe $\tan\beta$. 
For very large $\tan\beta (\gtrsim 100)$, the total width measurement is useful 
to constrain $\tan\beta$. 
\vspace{-1em}

\section{Summary}
\vspace{-1ex}

In this review, we have studied THDMs. 
The THDMs contain additional Higgs bosons, which provide rich phenomenology in the Higgs sector. 
These models are categorized by the Yukawa interactions under imposing a discrete symmetry. 
The experimental constraints from the flavor data and the LHC data on the THDMs are 
highly dependent on the type of the Yukawa interactions.  

In the Type-II THDM, the Yukawa interaction of non-standard Higgs bosons with the bottom quarks 
are enhanced for large $\tan\beta$. Such a parameter region is strongly constrained by the direct 
search for $H/A$ at the LHC and by $B$ decay data. Thus, small (large) $\tan\beta$ region together 
with light (heavy) additional Higgs bosons is experimentally allowed. 
In order to explore $\tan\beta$, three different methods, 
i) the branching ratio measurement of $H/A$, 
ii) the total width measurement of $H/A$, and 
iii) the branching ratio measurement of $h$, are applied. 
If the ILC energy is high enough to produce additional Higgs bosons, methods i) and ii) seems 
to be useful. If we observe deviations in the SM-like Higgs gauge interaction, the branching ratio 
measurement of the SM-like Higgs boson can probe $\tan\beta$ even when $H/A$ are heavy. 

In the Type-X THDM, the non-standard Higgs bosons are leptophilic, and hence it is difficult 
to be constrained. Therefore, additional Higgs bosons can be light enough to be produced at the ILC. 
The model can be tested both at the LHC (with high luminosity) and the ILC by looking at multi tau 
lepton signatures followed by $HA$ pair production. 
The mass determination of $H/A$ is also possible by the help of the collinear approximation. 
Similarly to the Type-II THDM, $\tan\beta$ can be probed by several methods.

\begin{acknowledgments}
\vspace{-2ex}
This article is partly based on a work in collaboration with S.~Kanemura 
 and H.~Yokoya which is in progress~\cite{Ref:KTY-TanB}. 
The work was supported, in part,
by the Grant-in-Aid for Scientific research from the Ministry of Education, Science, Sports, and
Culture (MEXT), Japan, No. 23104011.
\end{acknowledgments}
\vspace{-5ex}

\bigskip 

\end{document}